Nanoengineered Astronomical Optics

E.F. Borra[1], A.M. Ritcey[2], R. Bergamasco[1,2], P. Laird[1], J. Gingras[2], M. Dallaire[1], L. Da Silva[1], H. Yockell-Lelievre[2]

Centre d'Optique, Photonique et Lasers, Université Laval, Québec, Qc, Canada G1K 7P4

SUBJECT HEADINGS: telescopes , instrumentation: miscellaneous, instrumentation: adaptive optics

(SEND ALL EDITORIAL CORRESPONDANCE TO: E.F. BORRA)

RECEIVED_________________________________________

1 Département de Physique, Genie Physique et Optique

2 Département de Chimie

**ABSTRACT**

We describe a technology for the fabrication of inexpensive and versatile mirrors through the use of  a new type of nanoengineered optical material composed by the spreading of a self-assembling reflective colloidal film spread at the surface of a liquid. These new reflecting liquids offer  interesting possibilities for astronomical instrumentation. For example, they can replace mercury in conventional rotating liquid mirrors. The  main advantages offered include extremely low cost and, by coating a viscous liquid, the possibility of tilting the mirror by a few tens of degrees. We also have coated ferromagnetic liquids with these reflecting films. The resulting surfaces can be shaped by the application of a magnetic field, yielding reflecting surfaces that can have complicated shapes that can rapidly shift with time. These inexpensive and versatile optical elements could have numerous scientific and technological applications. Among possible astronomical applications, they could be used to make large inexpensive adaptive mirrors exhibiting strokes ranging from nanometers  to several millimeters.

## 1.  INTRODUCTION

Most astronomical optical instruments use glass optics for the primary mirror of the telescope as well as for auxiliary optical elements. Polished glass surfaces are expensive and it is difficult, if not impossible, to give them complex shapes. It is therefore worthwhile to find new technologies that would enable us to build inexpensive optics, of all sizes, capable of having highly aspheric active surfaces.

Liquid mirrors offer such a solution. They take advantage of the fact that the surface of a liquid is very smooth and takes the shape of an equipotential surface so that one readily gets an optical quality surface. For example, the fact that the surface of a liquid rotating in a gravitational field takes the shape of a parabola has been used to make inexpensive mirrors having excellent surface qualities. The technology is young but its performance is well documented by laboratory tests (Borra et al. 1992, Girard & Borra 1997, Tremblay & Borra 2000) as well as by observations (Sica et al. 1995, Hickson & Mulrooney 1998; Cabanac, Borra, & Beauchemin 1998). One also can make liquid mirrors by shaping the equipotential surfaces with force fields other than gravity. Magnetic fields, for example, can be used with ferromagnetic liquids.

While mercury is useful for conventional rotating liquid mirrors because of its reasonably good reflectivity and low melting temperature, its density is high so that the container and bearing have to be sturdy and relatively costly. Pure mercury is diamagnetic and cannot be easily deformed in a magnetic field. It can be rendered ferromagnetic, however it is difficult to obtain a stable mercury based ferromagnetic liquid. Furthermore, the high density of mercury necessitates large deforming forces and strong magnetic fields that require strong currents for the driving electromagnets, causing undue ohmic heating. It would therefore be useful to make magnetically shaped reflective liquids having better physical and chemical characteristics. We describe below such liquids based on colloidal surface films spread on liquid substrates.

## 2. DEPOSITION OF REFLECTIVE FILMS ON OIL-BASED SUBSTRATES

The mirrors reported in this paper are comprised of a thin reflective layer of silver nanoparticles supported on a liquid substrate. Stable interfacial assemblies of silver nano-particles have been previously reported in the literature and are known as Metal Liquid-Like Films, or MELLFs. (Yogev & Efrima1988; Gordon, McGarvey & Taylor1989). MELLFs were originally used in chemical experiments to study the chemicals that coated the colloidal grains.  We were intrigued by these colloidal systems that combine the optical properties of metals (reflectivity) with the fluidity of a liquid suspension and therefore seem well adapted to applications in the field of liquid optics. In an early paper (Borra, Ritcey, & Artigau 1999, hereafter referred to as paper I) we reported on early experiments. The preparation of a MELLF involves the fabrication of silver nanoparticles coated with an organic protective layer.  Experimental details have been previously described  (Yockell-Lelievre, Borra, Ritcey & Da Silva 2003, hereafter referred to as Paper II)).  In the present paper we report the successful subsequent spreading of the MELLF at the surface of an oil as described in the following procedure: An aqueous solution of an appropriate surfactant is first introduced into a container of desired dimensions.  For a rotating mirror, the motorised rotation of the container is initiated before the MELFF, mixed with paraffin oil (50% vol.), is slowly introduced onto the water surface.   Phase separation spontaneously occurs and a reflective film of silver nanoparticles forms at the oil surface.  The system typically becomes stable after about 45 minutes, at which time optical measurements can be carried out.

## 3.  NANOENGINEERED LIQUID MIRRORS

In Paper I we introduced nanoengineered mirrors that trapped colloidal particles at the interface between two liquids. In Paper II we discuss results obtained with water-based liquid mirrors where the colloidal particles are located at the top surface of water, a major improvement over Paper I where they were located at the interface between two liquids. In this section, we present results obtained with novel reflecting liquids made of silver coated oils. The ability to coat oil constitutes a major improvement of the technology, for the range of physical and chemical characteristics of the available liquids substrates is greatly extended. Furthermore, coating oil solves a major problem with water-base liquids: evaporation.

3.1 Interferometry

To develop the basic technology, we have mostly worked with flat mirrors since they are easier to handle and test interferometrically. Our tests are carried out with a Fizeau interferometer at a wavelength of 632.8 nm.

Fig. 1a  shows an interferogram obtained by measuring a 6.5 cm diameter sample of one of our dried MELLF surfaces spread on vegetable oil while Fig. 1b shows the corresponding wavefront obtained from the interferogram.  Both images indicate excellent optical quality. The resolution on the surface of the liquid was 0.1 cm. Deviations from a flat surface are mostly caused by defects in a flat glass folding mirror used to observe the liquid rather than defects in the sample mirror. There also is some contribution from waves on the liquid surface that are induced by  laboratory vibrations and from deformations at the edges due to surface tensions  near the rim of the relatively small container. The waves could be dampened by using thinner liquid layers (Borra et al. 1992) or more viscous liquid substrates. Even with these preventable defects, we can see that the statistics of the wavefront are excellent (RMS =  0.106 waves). Since the quoted RMS is on the wavefront, the mirror surface RMS deviations are $\lambda/20$.

Some minor problems with the coating technology must be noted. For example, large surfaces can be marred by a few bubbles, as well as dull spots caused by evaporating water droplets that leave solid residues. We are presently developing techniques to minimize these defects.

3.2 Reflectance curves

We measured the reflectivity curves of our samples as a function of wavelength with a spectrophotometer. The reflectivity curves for two water-based, an oil-based  mirror, as well as a coated ferrofluid are shown in Figure 2. The better reflectivities of the water samples presumably are due to the fact that, at the time of this writing, we have spent far more time working with water and have been able to find improved experimental conditions to maximize the reflectivity; while the oil coating technique was only discovered recently. The difference between the two water-based curves illustrates some of the progress achieved in a few months work following  Paper II.  Similar optimization of  the various experimental parameters affecting the reflectivity (such as the choice of surfactant and particle coating agent) has not yet been carried out on the oil based systems.  We are confident that we can improve significantly our reflectivity curves with additional work currently in progress.

We have made a 0.5-m diameter f/5 rotating mirror that uses a MELLF on liquid oil. We know from interferometry previously carried out on mercury and gallium liquid mirrors that liquid rotating surfaces are parabolical to a high accuracy so that we should expect that defects caused by the MELLF should have mostly high spatial frequencies. The knife-edge test, although not a very sophisticated one, is well-suited for our purpose since it can detect defects having very small amplitudes (as little as $1/1000^{th}$ of a wave in some cases) for defects having high spatial frequencies. A knife-edge test image of the 0.5-m mirror is shown in Figure 3. Comparing this image to previous knife-edge tests of liquid mirrors (Borra et al. 1985), we note the typical signature of a parabola, the absence of concentric ripples and the presence of small localized defects. We estimate that most defects have amplitudes $<\lambda/20$. The absence of ripples is due to damping in the more viscous liquid.

We have characterized the stability of these mirrors by measuring the variation of the reflectivity at 632.8 nm  with time over a period of a few weeks. We find that the reflectivity decreases by about 8 % over the first 10 days after a freshly made MELLF is deposited on oil surface. After that, the reflectivity stays constant for the next few weeks. We furthermore measured the variations of the full reflectivity curves over a one week period and found that the reflectivity decrease measured at 632.8 nm in fact only occurs at  wavelengths shorter than of about 800 nm.  No significant variation is observed  with time for longer wavelengths. Note that presumably, we can slow down the degradation of reflectivity with a proper choice of coating agent.

4.      Nanoengineered Ferromagnetic liquid mirrors .

Ferrofluids are multi-phase liquids in which ferromagnetic particles are held as a colloidal suspension in a carrier liquid. A wide range of  carrier liquids is available, the choice of carrier having a very strong impact on the hydrodynamic properties of the resulting ferrofluid. As a first approximation, the deformation of an inviscid magnetic liquid in a static magnetic field normal to the surface of the liquid is given by

$$\Delta h = B^2 (\mu_r - 1)/(2\mu_0 \rho g) \qquad (1)$$

Where $\rho$ is the density of the ferrofluid and B is the external magnetic field just above the ferrofluid surface. When the component of the magnetic field normal to the surface exceeds a critical value related to the interfacial tension and the density of the ferrofluid, a non-linear surface phenomenon instability, the Rosensweig instability, develops which renders the surface totally useless for our purpose. For typical light oil based ferrofluids the instability occurs around 84 Gauss, giving a maximum surface deformation of the order of 1 mm, well above the stroke requirement for adaptive optics.

We have built a ferrofluidic mirror having a diameter of 80 mm with an actuator spacing of 5.5 mm. The actuators consist of 5 mm diameter coils with soft ferrite cores that reduce current requirements by at least two orders of magnitude. Commands are generated using a desktop computer. The mirror surface was characterized with a Fizeau interferometer at a wavelength of 632.8 nm. Deformations of up to 10 μm have been measured; beyond which point the measurement becomes difficult due to the close fringe spacing. Qualitative measurements were also made of deformations on the order of several mm that were visible to the naked eye confirming the extremely large dynamic range of this type of mirror. The deformed surfaces appeared smooth to the unaided eye.

Fig 4 shows an interferogram and the resulting wavefront produced using the 25 central actuators in the mirror. We only have generated a piston term because the number of actuators is too small to show higher terms. We are now upgrading our system to 100 actuators and developing algorithms to generate Zernicke polynomials. The central plateau has an RMS wavefront deviation from flatness of 0.06 waves. Improved surface quality should be possible but will require improvements to our control software. This is another reason why we only try to demonstrate the piston term at this time.

The response time of these mirrors depends on the size of the deformation and the viscosity of the carrier liquid. Efforts are currently underway to characterize these parameters more precisely. Early indications are that the mirror frequency will exceed 100 Hz for deformations on the order of 10 μm for typical light oil based fluids.

The influence function of a liquid deformable mirror, which is the deformation measured at an actuator due to an applied current in an adjacent coil, has a coupling in the range of 18-33% which is consistent with conventional deformable mirrors. Individual coils produce a Gaussian influence function however the vector nature of the magnetic fields generated by the coils produces small dips between actuators. Proper selection of ferrofluid depth and coating film smoothes out these dips. More work remains to optimize actuator geometry and ferrofluid thickness to further improve surface quality.

Power dissipation in the actuators of a deformable mirror is a common problem because astronomical optics must preserve seeing quality. The use of high magnetic susceptibility ferrofluids with low density allows operation at low coil currents. A 10 μm deformation of a 2mm thick ferrofluid requires a power dissipation of approximately 8 mW per coil. At this level, no active cooling was required at room temperature. Additionally, because the coils are low resistance and the currents required are quite small, no high voltage stage is required, unlike for piezoelectric actuators.

## 4.  DISCUSSION AND CONCLUSION.

This article reports on optical tests of mirrors made with a chemical technique that allows for the coating of the air-liquid interface of water or oil based substrates with a reflective layer of self-assembling metallic nanoparticles. This is a significant improvement over the work in paper I, where the reflecting layer was trapped at the interface between two liquids.  In the new configuration, the reflecting layer coats the first surface seen by incident light.  The ability to coating oil, as opposed to water, offers significant advantages since it increases the number of available liquids and solves the evaporation problem of water-based liquids.

Interferometry shows that the coated surfaces have excellent optical qualities. Reflectivity measurements show peak values of the order of 80%, for water based MELLFs which is less than the reflectivity of silver or aluminum coated glass but usable nevertheless. The reflectivity of oil-based MELLF is lower but this is presumably due to the fact that, being newer, we have not done as much work with them. Consider that, in practice, glass-coated mirrors are seldom recoated, hence degrade with time as they get covered with dust and dirt. This is particularly true for telescope mirrors used in astronomy, the space and atmospheric sciences which are exposed to the elements. Our reflecting layers are inexpensive and easy to generate so that they can be refreshed often. We are confident that the reflectivities can be improved with additional work currently in progress. Knife-edge tests of a 0.5-m diameter rotating mirror indicates a reasonable surface quality. We are currently working to improve the deposition technique to further improve surface quality and reflectivity.

We expect that nanoengineered liquid mirrors should be useful for scientific and engineering applications. The technology is interesting for large optics, such as large parabolic mirrors, because of its extremely low cost. Because the mirror basically is made of a 1-mm thick layer of liquid having the density of oil, a 10-m diameter coated liquid mirror would weigh less than 100 kg. The costs of the mirror and supporting hardware, which depend on the weight of the reflecting liquid, should be quite low , probably of the order of a few tens of thousands of dollars. The quantities of chemicals needed are very small and the coating process itself is not labor demanding. This would usher in an era of inexpensive large telescopes readily available  to astronomers. Making a 4-m class liquid mirror necessitates few resources and can easily be made in a University environment with mostly student labor: We have done it. Note also that, while conventional mercury mirrors can only  be tilted by a few arcseconds, thus limiting their usefulness, rotating mirrors that use viscous liquids could  be tilted by perhaps as much as a few tens of degrees (Paper I).

Coated ferromagnetic liquids are interesting because they can be used  to make versatile low-cost adaptive and active mirrors. Because their surfaces can be shaped with magnetic fields, one can generate complex, time varying surfaces difficult and expensive to make with conventional techniques. We have demonstrated a magnetically deformable ferromagnetic liquid mirror capable of producing strokes of more than 10 μm with relatively low power requirements.  The speed of this device for deformations up to 10 μm is expected to exceed 100 Hz. Further testing is planned to confirm this. These mirrors promise several advantages over adaptive mirrors presently available. Deformations of the order of a few millimeters are possible using this system although the power requirements and heat loads become significant. There should be essentially no limit in size so that active mirrors having diameters of the order of one meter having hundreds of thousands of actuators should be feasible. We estimate that the cost of the hardware should run at about 5 dollars /actuator so that mirrors with hundreds of thousands of actuators would cost of the order of a million dollars. Mass production of the actuators could presumably result in even lower costs. Meter-sized active mirrors with hundreds of thousands of actuators are needed for the new generation of extremely large telescopes.

Preliminary experiments indicate  that ferromagnetic  liquid mirrors can be used in a tilted or even inverted configuration with the addition of a suitable biasing magnetic field.  We have fabricated down-facing ferromagnetic liquid mirrors that appeared smooth to the unaided eye, but have not yet carried out optical testing. It may thus be possible to use a liquid mirror as an adaptive secondary located above the main mirror or in other interesting geometries. Because the surface of liquids can be deformed by a variety of other physical effects (e.g. electrical or thermal fields), nanoengineered optics constitutes a new class of versatile and inexpensive optical elements that could be used in a variety of astronomical instrumentation. We are beginning to investigate this technology.

This is a very young technology and there is a significant work to be done before it reaches the performance and robustness expected for routine scientific or industrial applications. The work carried out so far has been relatively simple and we have reached with ease several milestones with the relatively modest resources typically available in a Canadian University, boding well for the future.

## ACKNOWLEDGMENTS

This research was supported by the Natural Sciences and Engineering Research Council and The Canadian Institute for Photonics Innovation. L.Vieira da Silva Jr. wishes to thank the Conselho Nacional de Desenvolvimento Científico e Tecnológico, CNPq , Brazil for their financial support.

## REFERENCES .

**FIGURE CAPTIONS.**

Fig. 1a: It shows an interferogram obtained by measuring a 6.5 cm diameter sample of an oil surface coated with a MELLF. The resolution on the surface of the liquid was 0.1 cm. Figure 1b shows the wavefront obtained from that interferogram. They show an excellent optical quality, as evidenced by the 0.1 waves RMS on the wavefront ($\lambda/20$ on the surface).

Fig. 2: It gives the reflectivity curves for two water-based and an oil-based mirror, as well as a coated ferrofluid. The difference between the two water-based curves illustrates some of the progress achieved in a few months work following Paper II. The better reflectivity curve of the water based mirror is probably due to the fact that we have spent far more time working with water-based liquids.

Figure 3: It shows the image obtained from a knife-edge test carried out on a 0.5-m f/5 rotating parabolic liquid mirror composed of a MELLF spread on paraffin oil. We can clearly see the signature of a parabola. The defects seen in the figure are discussed in the text.

Fig 4: It shows a piston term wavefront produced using the 25 central actuators in the ferrofluid mirror as well as the interferogram of the entire mirror. This wavefront was obtained on an uncoated ferrofluid.

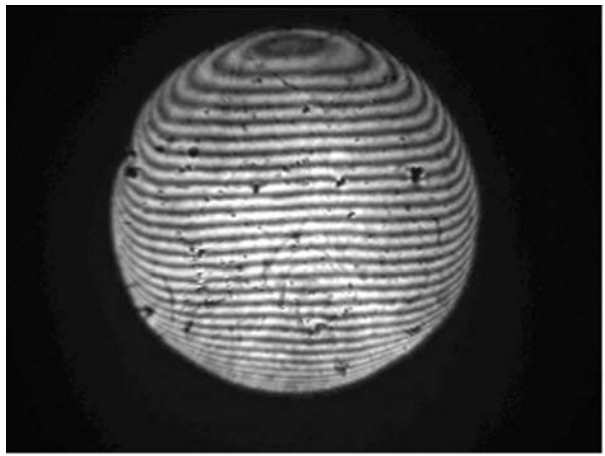

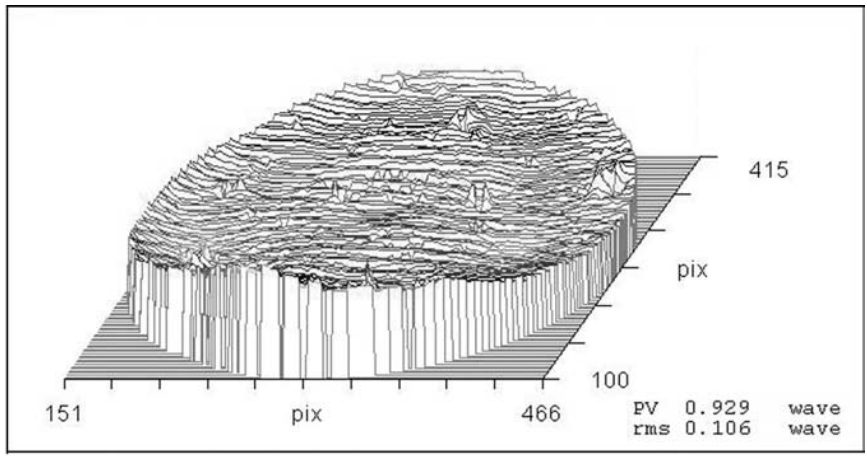

1a

1b

Figure 2

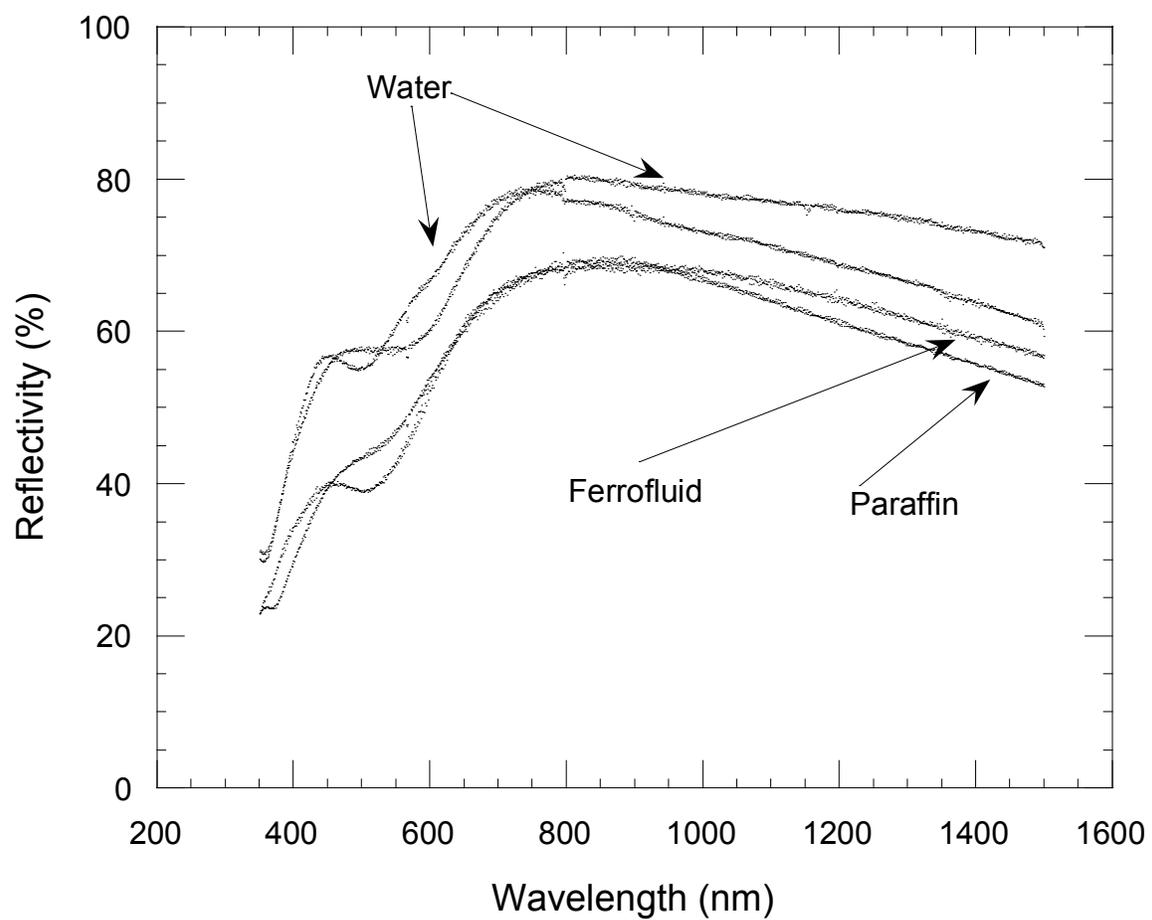

Figure 3

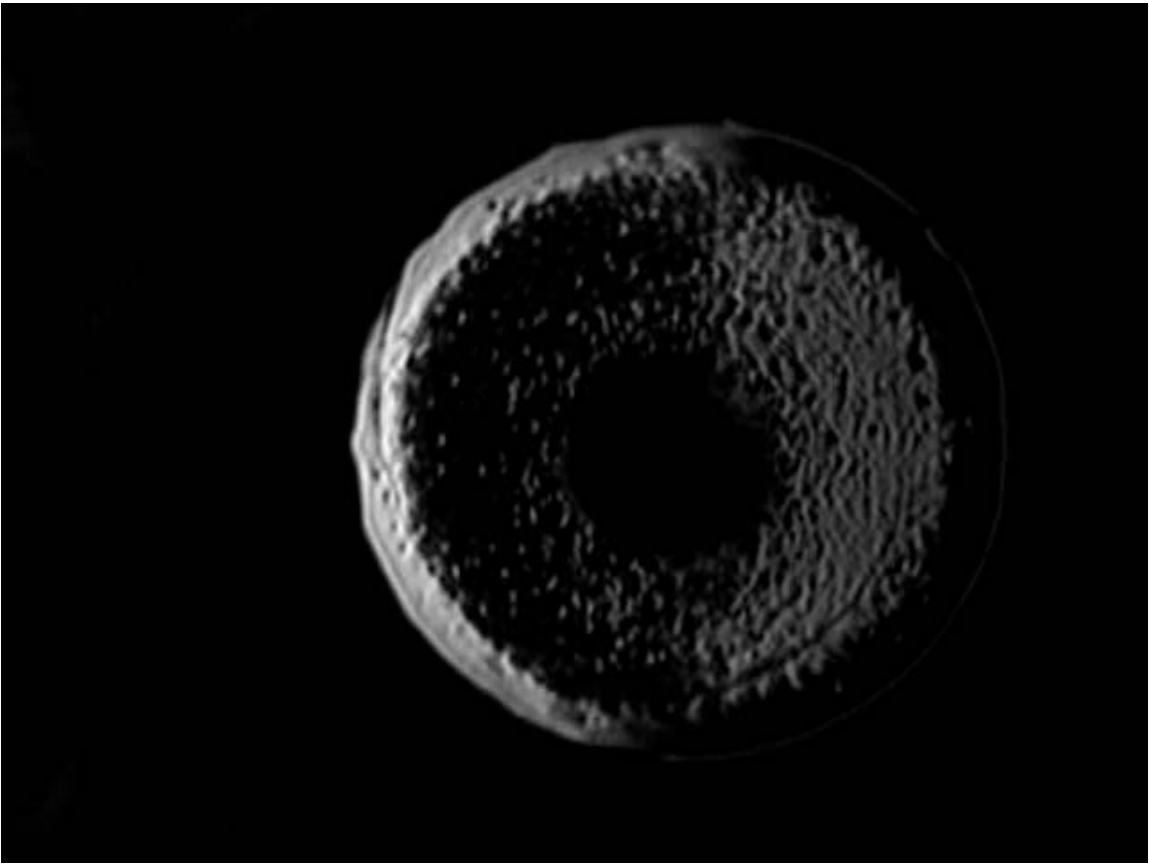

Figure 4

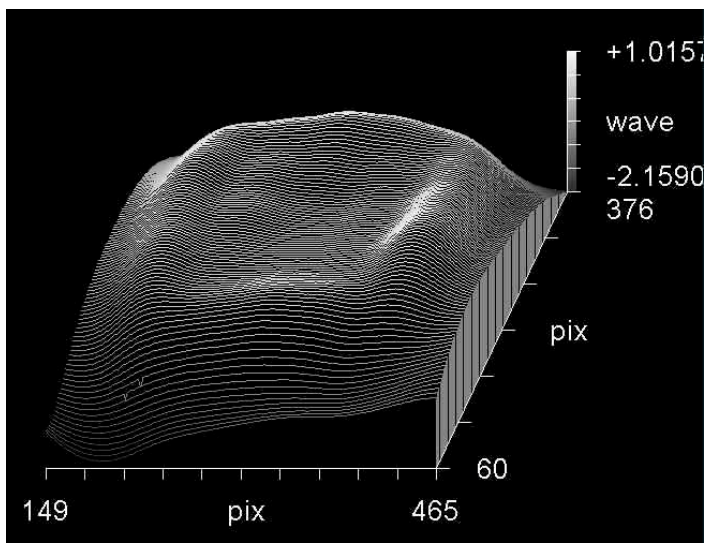
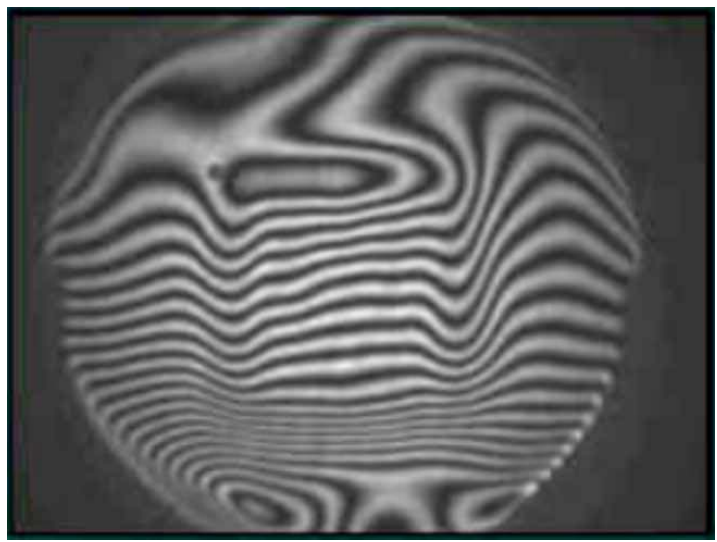